\documentclass[secnumarabic, nobibnotes,  nofootinbib, aps,
prd]{revtex4}%

\usepackage{amsmath}
\usepackage{amssymb}
\usepackage{graphicx}

\begin{document}
\title{The QCD evolution of TMD in the covariant approach
\footnote{To appear in the proceedings of the XVI Workshop on High Energy Spin Physics DSPIN-15, Dubna, Russia, September 8-12, 2015}}

\author{A.V. Efremov$^{a}$, O.V. Teryaev$^{a}$ and \underline{P. Zavada}$^{b}$}
\affiliation{$^{a}$Bogoliubov Laboratory of Theoretical Physics, JINR, 141980 Dubna, Russia}
\affiliation{$^{b}$Institute of Physics AS CR, Na Slovance 2, CZ-182 21 Prague 8, Czech Republic}

\begin{abstract}
The procedure for calculation of the QCD evolution of transverse momentum dependent distributions within the covariant approach is suggested. The standard collinear QCD evolution together with the requirements of relativistic invariance and rotational symmetry of the nucleon in its rest frame represent the basic ingredients of our approach. The obtained results are compared with the predictions of some other approaches.
\end{abstract}
\maketitle

\section{Introduction}
In our previous study we discussed various aspects of the covariant
quark-parton model, see \cite{Zavada:2013ola,Efremov:2010mt,
Efremov:2009ze,Zavada:2009ska,Zavada:2007ww,Zavada:2011cv} and citations
therein. The discussion included distribution and structure functions
(PDF-parton distribution function, TMD-transverse momentum distributions,
unpolarized and polarized structure functions) in the leading order. For a
fixed $Q^{2}$ we obtained the set of relations and rules, which interrelate
some of them, for example PDF withTMD:
\begin{equation}
q\left(  x\right)  \rightleftarrows q\left(  x,p_{T}\right)  , \label{ei1}%
\end{equation}
as we have studied in \cite{Efremov:2010mt, Efremov:2009ze}.

The aim of the present talk is to discuss a possible procedure for the
$Q^{2}-$ evolution in the covariant approach. The starting point will be the
use of a standard algorithm for the collinear QCD evolution of integrated
distribution: $q\left(  x\right)  \rightarrow$ $q\left(  x,Q^{2}\right)  $.
Then, the evolved distribution $q\left(  x,Q^{2}\right)  $ is inserted into
the original relation (\ref{ei1}). In this way we have defined the procedure
for the evolution of TMD:
\begin{equation}
q\left(  x\right)  \rightarrow q\left(  x,Q^{2}\right)  \rightarrow q\left(
x,p_{T},Q^{2}\right)  . \label{ei2}%
\end{equation}
In the present study we will discuss unpolarized distributions only. In the
next section we collect corresponding relations defining the procedure
(\ref{ei1}) in detail. The last section involves the numerical results and its discussion.

\section{Evolution of TMD}

The well known algorithm for evolution of the unpolarized PDF reads:%
\begin{equation}
\frac{d}{d\ln Q^{2}}q\left(  x,Q^{2}\right)  =\int_{x}^{1}\frac{dy}{y}P\left(
\frac{x}{y}\right)  q\left(  x,Q^{2}\right)  , \label{e1}%
\end{equation}
where $P$ is the corresponding splitting function. We have shown
\cite{Zavada:2007ww} that in the leading order approach the relativistic
invariance and spherical symmetry in the nucleon rest frame imply%
\begin{equation}
\rho_{q}(p,Q^{2})=4\pi p^{2}MG_{q}(p,Q^{2})=-x^{2}\left(  \frac{q(x,Q^{2})}%
{x}\right)  ^{\prime}=q\left(  x,Q^{2}\right)  -xq^{\prime}\left(
x,Q^{2}\right)  ;\qquad p\left(  x\right)  =\frac{Mx}{2}. \label{e2}%
\end{equation}
where $\rho_{q}(p)$ is the probability distribution of the quark momentum
$p=\left\vert \mathbf{p}\right\vert $ in the rest frame, $f^{\prime}\left(
x,Q^{2}\right)  $ denotes $df/dx.$ This relation is valid for any sufficiently
large $Q^{2},$ when the quark can be considered effectively free in any frame
in the sense of Ref. \cite{Zavada:2013ola}. The relation (\ref{e1})\ implies%
\begin{equation}
\frac{d}{d\ln Q^{2}}q^{\prime}\left(  x,Q^{2}\right)  =-\frac{1}{x}P\left(
\frac{x}{x}\right)  q\left(  x,Q^{2}\right)  +\int_{x}^{1}\frac{dy}{y}\frac
{d}{dx}P\left(  \frac{x}{y}\right)  q\left(  y,Q^{2}\right)  . \label{b1}%
\end{equation}
Since%
\begin{equation}
\frac{d}{dx}P\left(  \frac{x}{y}\right)  =-\frac{y}{x}\frac{d}{dy}P\left(
\frac{x}{y}\right)  , \label{b2}%
\end{equation}
integration by parts gives%
\begin{equation}
\frac{d}{d\ln Q^{2}}\left(  xq^{\prime}\left(  x,Q^{2}\right)  \right)
=\int_{x}^{1}\frac{dy}{y}\frac{d}{dx}P\left(  \frac{x}{y}\right)  \left(
yq^{\prime}\left(  y,Q^{2}\right)  \right)  . \label{b3}%
\end{equation}
This equality together with Eqs. (\ref{e1})\ and (\ref{e2})\ gives\ %

\begin{equation}
\frac{d}{d\ln Q^{2}}\rho_{q}\left(  p\left(  x\right)  ,Q^{2}\right)
=\int_{x}^{1}\frac{dy}{y}P\left(  \frac{x}{y}\right)  \rho_{q}\left(  p\left(
y\right)  ,Q^{2}\right)  \label{b4}%
\end{equation}
or equivalently%
\begin{equation}
\frac{d}{d\ln Q^{2}}G_{q}\left(  p,Q^{2}\right)  =\frac{1}{4\pi p^{2}M}%
\int_{p}^{M/2}\frac{dp%
\acute{}%
}{p%
\acute{}%
}P\left(  \frac{p}{p%
\acute{}%
}\right)  \rho_{q}\left(  p%
\acute{}%
,Q^{2}\right)  . \label{e3a}%
\end{equation}

Further, in Ref. \cite{Zavada:2009ska,Zavada:2007ww} (see also
\cite{D'Alesio:2009kv}) we proved the relation for the unpolarized TMD:%
\begin{equation}
f_{1}^{q}(x,p_{T},Q^{2})=MG_{q}(\tilde{p},Q^{2})=\frac{1}{4\pi\tilde{p}^{2}%
}\left(  q\left(  \xi,Q^{2}\right)  -\xi q^{\prime}\left(  \xi,Q^{2}\right)
\right)  , \label{e4}%
\end{equation}
where%
\begin{equation}
\tilde{p}(x,p_{T})=\frac{M\xi}{2},\qquad\xi=x\left(  1+\left(  \frac{p_{T}%
}{Mx}\right)  ^{2}\right)  . \label{e6}%
\end{equation}
Relation (\ref{e4}) with the use of (\ref{e3a}) give the TMD evolution:%
\begin{equation}
\frac{d}{d\ln Q^{2}}f_{1}^{q}(x,p_{T},Q^{2})=\frac{1}{4\pi\tilde{p}^{2}}%
\int_{\tilde{p}}^{M/2}\frac{dp%
\acute{}%
}{p%
\acute{}%
}P\left(  \frac{\tilde{p}}{p%
\acute{}%
}\right)  \rho_{q}\left(  p%
\acute{}%
,Q^{2}\right)  \label{e5}%
\end{equation}
Note the same splitting function in convolutions (\ref{e1}), (\ref{b4}),
(\ref{e3a}) and (\ref{e5}), which follows from the fact the distributions
$q\left(  x,Q^{2}\right),$ $\rho_{q}(p,Q^{2})$ and $f_{1}^{q}(x,p_{T}%
,Q^{2})$ are equivalent, as a result of relativistic invariance and rotational symmetry.

At the same time, in an accordance with Eq. (\ref{e4}), the evoluted TMD can
be expressed directly by means of the evoluted PDF. In the next this relation
will be used for numerical calculations of TMD evolution.

\begin{figure}[t]
\includegraphics[width=15cm]{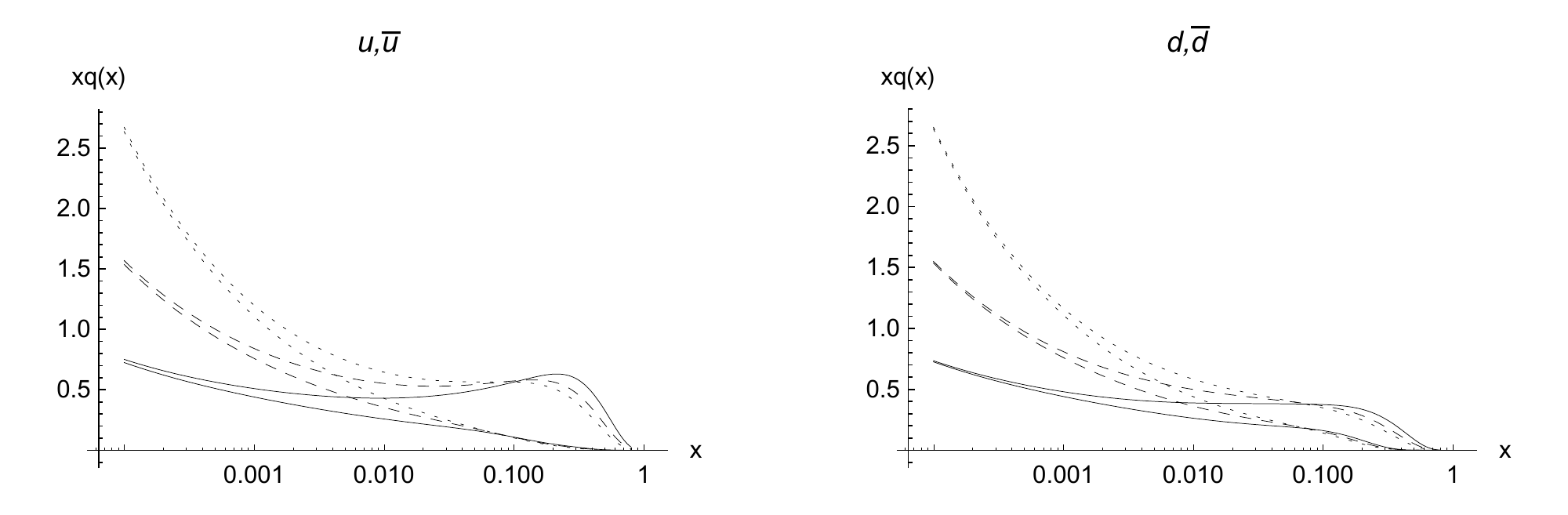}\caption{Input PDF of $u,\bar{u}$
(left) and $d,\bar{d}$ (right) quarks at different scales: $Q^{2}=4,40,400GeV$
(solid, dashed, dotted curves)}%
\label{fig1}%
\end{figure}\begin{figure}[t]
\includegraphics[width=15cm]{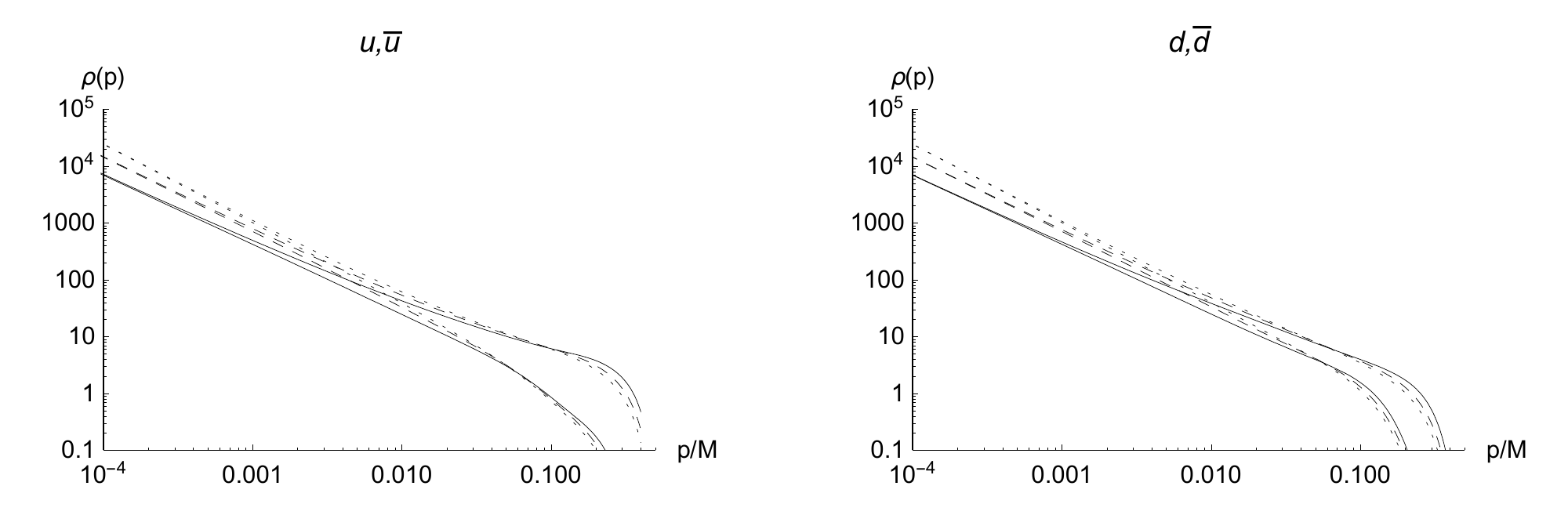}\caption{Distribution of the momentum
in the nucleon rest frame for the quarks $u,\bar{u}$ (left) and $d,\bar{d}$
(right) at different scales: $Q^{2}=4,40,400GeV$ (solid, dashed, dotted
curves)}%
\label{fig2}%
\end{figure}

\section{Results}

\begin{figure}[t]
\includegraphics[width=15cm]{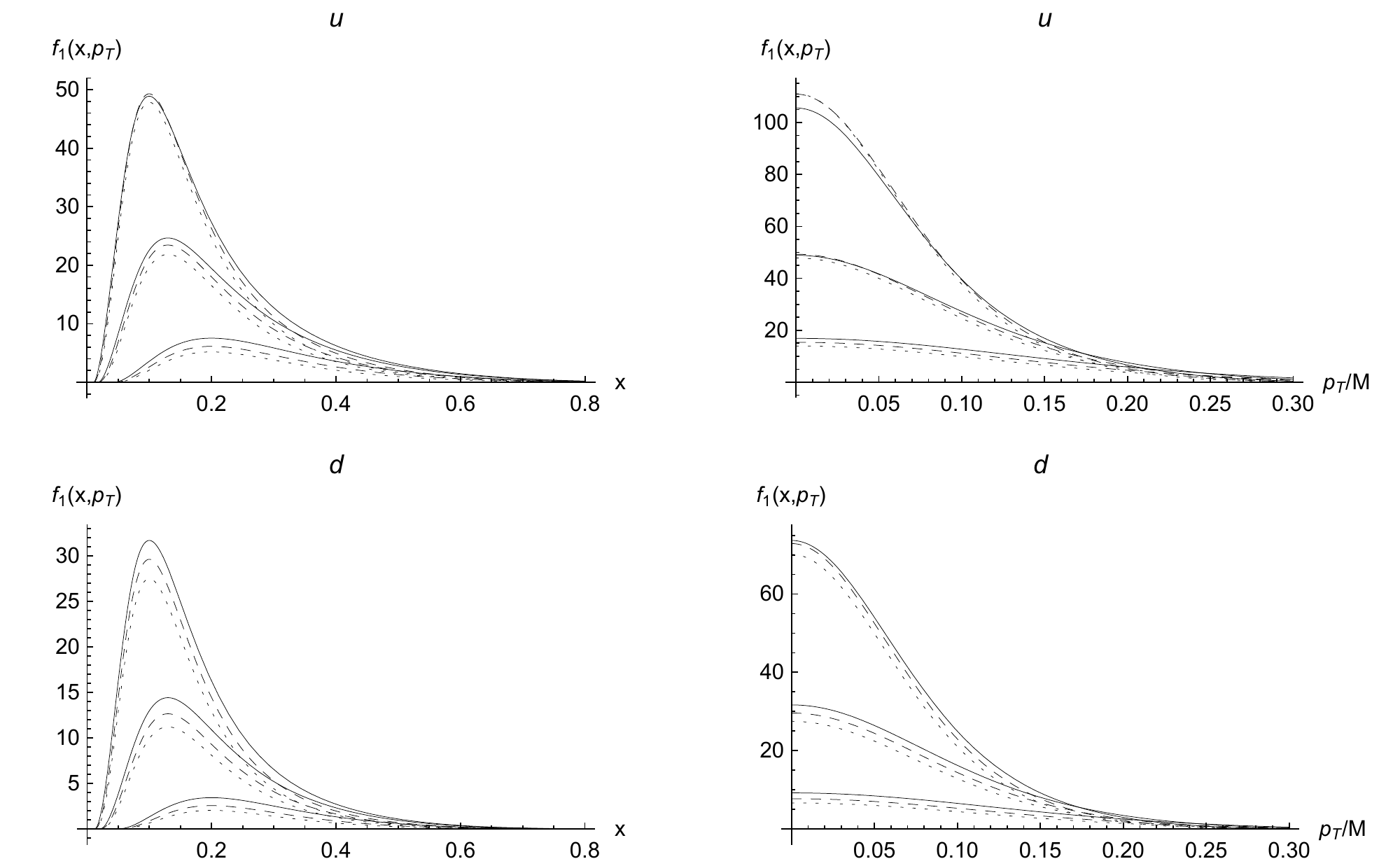}\caption{TMD at different scales:
$Q^{2}=4,40,400GeV$ (solid, dashed, dotted curves) for $u$ and $d$ quarks.
Sets of curves in left panels (from top) correspond to fixed $p_{T}%
/M=0.1,0.13,0.20$. The curves in left panels (from top) correspod to fixed
$x=0.18,0.22,0.30$.}%
\label{fg3}%
\end{figure}

For numerical calculation we have used the PDF set MSTW2008(LO)
\cite{Martin:2009iq} at the three different scales, which are displayed in
Fig.\ref{fig1}. In the next figure (Fig.\ref{fig2}) we have shown their
representation in terms of the distributions $\rho_{q}(p,Q^{2})$ calculated
from the relation (\ref{e2}). Finally, in Fig.\ref{fg3} we have displayed the
corresponding TMDs. One can observe two important features:

i) The transverse moments of quarks satisfy the condition $p_{T}<M/2$. This
condition follows from the constraint $0<x_{B}<1$ and the condition of
relativistic invariance and rotational symmetry as we have shown in Ref.
\cite{Zavada:2011cv}.

ii) Dependence of TMD on the scale $Q^{2}$ is rather weak.

These results are well compatible with the predictions obtained within the
statistical approach and presented in Ref. \cite{Bourrely:2013yti}. On the
other hand our results on TMD differs rather substantially e.g. from the
results of the QCD evolution in Ref. \cite{Aybat:2011zv}. The possible
resolution of this contradiction is that evolution considered here may be
attributed to the 'soft' non-perturbative component of TMDs and even can be
used to disentangle it from the 'hard' perturbative component. Apparently these
discrepancies require further study.

\newpage

\begin{acknowledgments}

Authors of this work were supported by the Votruba-Blokhitsev Program of JINR
Dubna.  P.Z. was supported also by the project LG130131 of the MEYS (Czech Republic).
\end{acknowledgments}

\end{document}